\documentstyle[pra,aps]{revtex}

\begin{document}
\draft
\title{The Classical Limit of Quantum Mechanics and the Fej\'{e}r Sum of the
Fourier Series Expansion of a Classical Quantity}
\author{Quan-Hui Liu}
\address{Department of Physics, Hunan University, Changsha, 410082, China}
\address{and}
\address{Institute of Theoretical Physics, Chinese Academy of Science, P.O. Box 2735,%
\\
Beijing\\
100080, China\thanks{%
Mailing address. E-mail address: liuqh@itp.ac.cn}}
\date{\today}
\maketitle

\begin{abstract}
In quantum mechanics, the expectation value of a quantity on a quantum
state, provided that the state itself gives in the classical limit a motion
of a particle in a definite path, in classical limit goes over to Fourier
series form of the classical quantity. Different from this widely accepted
point of view, a rigorous calculation shows that the expectation value on
such a state in classical limit exactly gives the Fej\'{e}r's arithmetic
mean of the partial sums of the Fourier series.
\end{abstract}

\pacs{03.65.Ca, 03.65.Sq}

\section{Introduction}

It is widely accepted that the expectation value of a quantity in any
quantum state must become, in the classical limit, simply the classical
value of the quantity, provided that the state itself gives, in the limit, a
motion of a particle in a definite path\cite{lan}. And the expectation value
in the classical limit gives the Fourier series form of the classical
quantity\cite{lan}. In this Letter, we would like to point out that the
expectation value on such a state in classical limit is exactly the
Fej\'{e}r's arithmetic mean of the partial sums of the Fourier series. The
Fourier series itself and the Fej\'{e}r's arithmetic mean of the partial
sums of the Fourier series are different. Even through both of them can be
used to represent a periodic function, conceptually they are totally
different from each other. Furthermore, the former is worse than the latter
in convergence\cite{cou}.

In our approach, the classical limit will refer to the following
mathematically well-established one\cite{has}: 
\begin{equation}
n\rightarrow \infty ,\hbar \rightarrow 0,n\hbar
=an~~appropriate~~classical~~action.
\end{equation}

In order to obtain a definite classical path in classical limit, we must
start from a wave function of a particular form\cite{lan}. A routine way to
construct such a wave function is $\sum_n c_n \psi_n $, where the
coefficients $c_n$ are noticeably different from zero only in some range $%
\delta n$ of values of the quantum number $n$ such that $1<<\delta n<< n$;
the numbers $n$ are supposed large and the superposed states $\psi_n$ have
nearly the same energy. This particular wave function, commonly called wave
packet, suffices to discuss the classical limit of quantum mechanics. To
note that the choice of a set of coefficients $c_n$ is a matter of
convenience. The only requirement on the distributions of $c_n$ among $n$ is
that they must be equal to each other in classical limit, otherwise we would
give results rather than correct classical mechanical ones (we will come
back to this point in Section IV). For example, the commonly used Gaussian
distribution or Poisson distribution meets this requirement in classical
limit, for both give the same thing: the distribution of $c_n$ among $n$
being approximately equal. So, the characteristic of the classical limit of
quantum mechanics is involved in the classical limit of the following wave
packet, which is a linear combination of energy eigenfunctions of large
quantum number with equal weight (equally-weighted wave packet, for
abbreviation) of the following form:

\begin{equation}
|\psi (t)>=\frac 1{\sqrt{2N+1}}\sum\limits_{m=-N}^N|n+m>exp(-iE_{n+m}t/\hbar
),
\end{equation}
where $n$ and $N$ are positive integers and $N>0$, $n-N>0$. In fact, this
wave packet not only does the job well but also is very easy to handle. As
we will see later, for matching with the exact classical result in classical
limit, $n$ and the parameter $N$ are necessarily large in physics, or
approach infinity in terms of mathematics. It is what we expected. However,
in the following section II, the expectation value on the equally-weighted
wave packet in the classical limit will be shown to be the Fej\'{e}r's
arithmetic mean of the partial sums of the Fourier series, rather than the
Fourier series itself. In section III, an example will be given. In section
IV, a brief discussion and conclusion will be presented.

\section{Classical limit of equally-weighted wave packet}

The expectation of a physical observable $f$ on the wave packet is 
\begin{equation}
<\psi (t)|f|\psi (t)>=\frac 1{2N+1}\sum\limits_{m^{\prime
}=-N}^N\sum\limits_{m=-N}^N<n+m^{\prime }|f|n+m>exp[i(E_{n+m^{\prime
}}-E_{n+m})t/\hbar].
\end{equation}

For the simple and integrable quantum systems having classical
correspondence, the Bohr's correspondence principle holds true. It asserts
that in classical limit $(E_{n+m^{\prime }}-E_{n+m})/\hbar =(m^{\prime
}-m)\omega $, with $\omega $ denoting the classical frequency\cite{lan}.
Also in the classical limit, the matrix element $<n+m^{\prime
}|f|n+m>=f_{m^{\prime }-m}$ is the $(m\prime -m)th$ Fourier component of the
corresponding classical quantity $f(t)$ in terms of the ordinary Fourier
series\cite{gree}. The latter relation appears trivial to some, Landau and
Lifshitz for instance\cite{lan}, and is quite new to others. It should be
emphasized that the two relations above are generally applicable only in the
classical limit \cite{gree}. One should not confuse the Bohr's
correspondence principle, $(E_{n+m^{\prime }}-E_{n+m})/\hbar =(m^{\prime
}-m)\omega $, with the difference of energy eigenvalues for a harmonic
oscillator. One can easily verify that the Bohr's correspondence principle
is valid for hydrogen atom, rigid rotators and particles in an infinite
square-well potential, etc. Then we have: 
\begin{eqnarray}
<\psi (t)|f|\psi (t)> &=&\frac 1{2N+1}\sum\limits_{m^{\prime
}=-N}^N\sum\limits_{m=-N}^Nf_{m^{\prime }-m}exp[i(m-m^{\prime })\omega t] 
\nonumber \\
&=&\frac 1{2N+1}\sum\limits_{l=0}^{2N}\sum\limits_{s=-l}^{2N-l}f_s
exp[is\omega t]  \nonumber \\
&=&\frac 1{2N+1}\sum\limits_{l=0}^{2N}\Sigma (2N-l,-l),
\end{eqnarray}
where we have made variable transformations as 
\begin{equation}
s=m-m^{\prime },~~l=m+N,
\end{equation}
and used the symbol $\Sigma (\alpha ,\beta )$ which is defined by 
\begin{equation}
\Sigma (\alpha ,\beta )=\sum\limits_\beta ^\alpha f_s exp(is\omega t).
\end{equation}
Our aim is to show that the equation given by the last line of Eq.(4) and
the RHS of following Eq.(12) are identical although they look different from
each other. The following is a proof.

We study the sum in Eq.(4) and find that

\begin{equation}
\sum\limits_{l=0}^{2N}\Sigma (2N-l,-l)=\sum\limits_{l=0}^{N-1}\Sigma
(2N-l,-l)+\Sigma (N,-N)+\sum\limits_{l=N+1}^{2N}\Sigma (2N-l,-l).
\end{equation}
Breaking the term $\Sigma (2N-l,-l)$ in the first sum on RHS of above
equation into two parts as $\Sigma (l,-l)+\Sigma (2N-l,l+1)$, the RHS
becomes: 
\begin{equation}
\sum\limits_{l=0}^N\Sigma (l,-l)+\sum\limits_{l=0}^{N-1}\Sigma
(2N-l,l+1)+\sum\limits_{l=N+1}^{2N}\Sigma (2N-l,-l).
\end{equation}
The last term in Eq.(8) can be changed into the following form with
transformation $2N-l\rightarrow l$: 
\begin{equation}
\sum\limits_{l=N+1}^{2N}\Sigma (2N-l,-l)=\sum\limits_{l=0}^{N-1}\Sigma
(l,-2N+l).
\end{equation}
Then Eq.(8) becomes: 
\begin{eqnarray}
&&\sum\limits_{l=0}^N\Sigma (l,-l)+\sum\limits_{l=0}^{N-1}\Sigma
(2N-l,l+1)+\sum\limits_{l=0}^{N-1}\Sigma (l,-2N+l)  \nonumber \\
&=&\sum\limits_{l=0}^N\Sigma (l,-l)+\sum\limits_{l=0}^{N-1}\Sigma
(2N-l,-2N+l)  \nonumber \\
&=&\sum\limits_{l=0}^N\Sigma (l,-l)+\sum\limits_{l=N+1}^{2N}\Sigma (l,-l) 
\nonumber \\
&=&\sum\limits_{l=0}^{2N}\Sigma (l,-l),
\end{eqnarray}
where we have used the transformation $2N-l\rightarrow l$. Thus we finally
obtain: 
\begin{equation}
\sum\limits_{l=0}^{2N}\Sigma (2N-l,-l)=\sum\limits_{l=0}^{2N}\Sigma (l,-l).
\end{equation}

>From the definition of $\Sigma (\alpha ,\beta )$, Eq.(6), we immediately
know that $\Sigma (l,-l)$ is the $l$-th partial sum of the ordinary Fourier
series of the classical quantity $f(t)$. We know that in the classical limit
(1) together with limit $N\rightarrow \infty $, $\Sigma (N,-N)$ converges to
the classical quantity $f(t)$. In fact, according to the Fej\'{e}r's
summation theorem \cite{cou}, in the same limit, the following arithmetic
mean of the partial sums 
\begin{equation}
<\psi (t)|f|\psi (t)>=\frac 1{2N+1}\sum\limits_{l=0}^{2N}\Sigma (l,-l),
\end{equation}
converges uniformly to the classical quantity $f(t)$ provided the quantity $%
f(t)$ is continuous\cite{cou}. This means that for every observable $f$, the
expectation value in the equally-weighted wave packet in the classical limit
gives the corresponding classical quantity $f(t)$. This also means that, in
the same limit, the equally-weighted wave packet gives a motion of the
particle in a definite classical path. Thus, our proof is complete.

\section{Equally-weighted wave packet for a single one-dimensional harmonic
oscillator}

The wave packet is 
\begin{equation}
|\psi(t)>=\frac{1}{\sqrt{2N+1}}\sum\limits^{N}_{m=-N}|n+m>exp(-iE_{n+m}t/%
\hbar).
\end{equation}

We have the expectation values for quantities $H,H^{2},x,x^{2},p,p^{2}$ in
the following. 
\begin{eqnarray}
&<&\psi (t)|H|\psi (t)>=(n+\frac{1}{2})\hbar \omega . \\
&<&\psi (t)|H^{2}|\psi (t)>=[(n+\frac{1}{2})\hbar \omega ]^{2}+(n\hbar
\omega )^{2}\frac{N(N+1)}{3n^{2}}. \\
&<&\psi (t)|x|\psi (t)> \\
&=&(\frac{2}{2N+1}\sqrt{\frac{\hbar }{2\mu \omega }}\sum\limits_{m=-N+1}^{N}%
\sqrt{n+m})cos\omega t.  \nonumber \\
&<&\psi (t)|x^{2}|\psi (t)> \\
&=&(n+\frac{1}{2})(\frac{\hbar }{\mu \omega })+(\frac{\hbar }{\mu \omega })%
\frac{1}{2N+1}\sum\limits_{m=-N+2}^{N}\sqrt{(n+m)(n+m-1)}cos2\omega t. 
\nonumber \\
&<&\psi (t)|p|\psi (t)> \\
&=&(\frac{2}{2N+1}\sqrt{\frac{\hbar \mu \omega }{2}}\sum\limits_{m=-N+1}^{N}%
\sqrt{n+m})sin\omega t.  \nonumber \\
&<&\psi (t)|p^{2}|\psi (t)> \\
&=&(n+\frac{1}{2})\mu \hbar \omega -\mu \hbar \omega \frac{1}{2N+1}%
\sum\limits_{m=-N+2}^{N}\sqrt{(n+m)(n+m-1)}cos2\omega t.  \nonumber
\end{eqnarray}
It is easy to see that these results in the classical limit (1) in
conjunction with the limit 
\begin{equation}
N\rightarrow \infty ,~~N/n\rightarrow 0.
\end{equation}
are exactly the classical quantities. And they are respectively 
\begin{eqnarray}
<\psi (t)|H|\psi (t)> &=&E. \\
<\psi (t)|H^{2}|\psi (t)> &=&E^{2}. \\
<\psi (t)|x|\psi (t)> &=&\sqrt{\frac{2E}{\mu \omega ^{2}}}cos(\omega t). \\
<\psi (t)|p|\psi (t)> &=&\sqrt{2\mu E}sin(\omega t). \\
<\psi (t)|x^{2}|\psi (t)> &=&\frac{E}{\mu \omega ^{2}}+\frac{E}{\mu \omega
^{2}}cos(2\omega t) \\
&=&(<\psi (t)|x|\psi (t)>)^{2}.  \nonumber \\
<\psi (t)|p^{2}|\psi (t)> &=&\mu E-\mu E\cos (2\omega t) \\
&=&(<\psi (t)|p|\psi (t)>)^{2}.  \nonumber
\end{eqnarray}

\section{Discussion and Conclusion}

Before enclosing this Letter, the following points should be mentioned.

1. One should not confuse the classical state, the coherent state, Gaussian
or Poisson wave packet for instance, with the classical limit of quantum
mechanics. The Planck's constant $\hbar$ can not be treat as zero in
classical state of quantum mechanics while in classical mechanics it is
practically zero.

2. If one starts a wave packet of following from 
\begin{equation}
|\psi (t)>=\sum\limits_{m=-N}^N c_m|n+m>exp(-iE_{n+m}t/\hbar),
\end{equation}
where in classical limit both $n$ and the parameter $N$ are necessarily
large, but the coefficients $c_m$ do not be equally distributed among $m$
even in classical limit, the dependence of the expectation values $%
<\psi(t)|f|\psi(t)>$ on $n$ will be different. It means that, for a specific
system, the dependence of classical equation of motion on the classical
action would be different if the action is different. It is not the case in
classical mechanics. Thus, the characteristic of the classical limit of
quantum mechanics is indeed involved in the classical limit of the
equally-weighted wave packet.

3. The expectation value of a quantity in the equally-weighted wave packet
in the classical limit goes over to the Fej\'{e}r's arithmetic mean of the
partial sums of Fourier series form of the classical quantity, not the
Fourier series itself as widely accepted. The ordinary Fourier series
differs from its Fej\'{e}r sum in convergence, for some cases the former
does not converge, whereas the latter does \cite{cou}. In comparing with our
proof, the usual proof involves some approximations\cite{lan}, and is not
rigorous. We are confident that Fej\'{e}r's arithmetic mean of the partial
sums of Fourier decomposition of the classical quantity is the only possible
form representing a single classical orbit from classical limit of quantum
mechanics.

4. In our approach, we used a prerequisite that classical limit of the
matrix element $<n+m^{\prime}|f|n+m>=f_{m^{\prime }-m}$ is the $(m\prime
-m)th$ Fourier component of the corresponding classical quantity $f(t)$ in
terms of the ordinary Fourier series\cite{lan,gree}. In fact, this
prerequisite is not necessary. To note that the Fej\'{e}r sums can also be
used to represent a periodic function\cite{cou}, as does the usual Fourier
series, and we can then write the classical quantity directly in the form of
the Fej\'{e}r sums. Because the expectation value $<\psi (t)|f|\psi (t)>$,
Eq.(12), in classical limit gives nothing but a classical quantity, a
comparison of the Fej\'{e}r sums form of the classical quantity with the
classical limit of $<\psi (t)|f|\psi (t)>$ given by Eq.(12) directly leads
to the prerequisite.

5. Our study implys that one may study that physical significance of a
constructed quantity $\sum_m f_{mn} exp[i(E_m-E_n)t/\hbar]$ or its like for
it in classical limit corresponds to the physical quantity in terms of
ordinary Fourier series. In fact there is already such a theory\cite{huang}.

6. Our approach is straightforward and simple, but the results appear exact
and new. Obviously, our results do not support the assertion that pure-state
quantum mechanics in classcall limit only reduces to classical statistical
mechanics then nothing more to it\cite{ball1,ball2,anj}.

\section{Acknowledgment}

I am indebted to Profs. Huan-Wu Peng, Ou-Yang Zhong-Can and Zu-Sen Zhao for
enlightening discussions, to Dr. Zhou Haijun for useful comments and to
Prof. V. Srivastava for critical reading of the manuscript. This subject is
supported by Grant No: KJ952-J1-404 of Chinese Academy of Science.

\end{document}